\documentstyle[12pt]{article}
\input epsf
\begin{document}

\rightline{RU-94-35}
\rightline{hep-ph/9407401}
\baselineskip=18pt
\vskip 0.5in
\begin{center}
{\bf \LARGE Experiments to Find or Exclude \\ a Long-Lived, Light
Gluino\\ }
\vspace*{0.7in}
{\large Glennys R. Farrar}\footnote{Research supported in part by
NSF-PHY-91-21039} \\
\vspace{.05in}
{\it Department of Physics and Astronomy \\ Rutgers University,
Piscataway, NJ 08855, USA}

\end{center}
\vskip  0.5in

{\bf Abstract:}
Gluinos in the mass range $\sim 1\frac{1}{2}-3\frac{1}{2}$ GeV
are absolutely excluded.  Lighter gluinos are allowed, except for
certain ranges of lifetime.  Only small parts of the
mass-lifetime parameter space are excluded for larger masses unless
the lifetime is shorter than $\sim 2 \times 10^{-11} \left(
\frac{m_{\tilde{g}}}{1 {\rm GeV}}\right) $ sec.  Refined mass and
lifetime estimates for $R$-hadrons are given, present direct and
indirect experimental constraints are reviewed, and experiments
to find or definitively exclude these possibilities are suggested.

\thispagestyle{empty}
\newpage
\addtocounter{page}{-1}

\section{Introduction}
\label{intro}
\hspace*{2em}

Short-lived gluinos could be defined to be those which decay before
interacting hadronically in a detector or beam-dump.  Their decay
produces the lightest neutralino (lsp)\footnote{Generally, a
superposition of the SUSY partners of the photon, $Z^0$ and neutral
Higgses, not considering effects of a possible light gravitino.} which
escapes the detector or dump without interacting, carrying with it much
of the gluino's energy and momentum.  Short-lived gluinos are excluded
for masses $\,\raisebox{-0.13cm}{$\stackrel{\textstyle<}
{\textstyle\sim}$}\, 160$ GeV\cite{cdf:gluinolim2} by the absence of
characteristic missing energy events in the FNAL collider.  Thus to be
lighter than this, gluinos must be long-lived\footnote{The long-lived
gluino window was first pointed out in ref. \cite{f:51}.  It was
subsequently discussed in the work of ref. \cite{deq}.}.  It is
natural for gluinos to be much lighter than squarks if their masses
are entirely radiative in origin.  In that case, if the SUSY and
elecroweak symmetry breaking scales are less than $\sim10$ TeV, gluino
and lsp masses will range from of order 100 MeV to of order 30
GeV\cite{bgm,f:96}.  This is the mass range explored here.

A gluino in the mass range $\sim 1.5 - 3.5$ GeV is excluded, whatever
its lifetime, from the absence of a peak in the photon energy spectrum
in radiative Upsilon decay.  This is because two gluinos with mass in
that range would form a pseudoscalar bound state, the
$\eta_{\tilde{g}}$, whose branching fraction in $\Upsilon \rightarrow
\gamma \eta_{\tilde{g}}$ can be reliably computed using perturbative
QCD and is predicted\cite{keung_khare,kuhn_ono,goldman_haber} to be
greater than the experimental upper
bound\cite{tutsmunich,cusb}\footnote{The range excluded by the CUSB
experiment is incorrectly claimed to extend to lower gluino masses, by
using the pQCD results of refs.
\cite{keung_khare,kuhn_ono,goldman_haber} out of their range of
validity.  A detailed analysis of the actual excluded range in given
in ref. \cite{f:93}.  The lower limit for validity of a pQCD,
non-relativistic potential model description of an $\eta_{\tilde{g}}$ was
taken to be $\sim 3$ GeV, mainly by analogy with the success of the
same description of charmonium.  However since the effective value of
the coupling is so much stronger due to the larger color charge of the
gluino in comparison to a quark, even a 3 GeV $\eta_{\tilde{g}}$ may not be
in the perturbative regime, in which case the range of validity of the
CUSB procedure may not be even this large.  Note that any gluino whose
lifetime is longer than the strong interaction disintegration time of
the $\eta_{\tilde{g}}$, i.e., $\tau
\,\raisebox{-0.13cm}{$\stackrel{\textstyle>}
{\textstyle\sim}$}\, \sim 10^{-22}$ sec, will produce
the requisite bump in the photon energy spectrum, and thus be
excluded by CUSB.}.

In this paper I address the question of whether long-lived gluinos
having mass less than $\sim 1.5$ or greater than $3.5$ GeV are
excluded on other grounds.  Many experiments which are commonly cited
as ruling out gluinos of this mass range actually provide only weak
limits when one takes account of the gluino lifetime.  These
experiments as well as the most powerful indirect constraints, which
are also presently unable to exclude this mass range, will be reviewed
below.  My purpose here is to propose tests which will {\it
unambiguously} demonstrate or exclude the existance of light gluinos.

An inevitable consequence of the existance of a long-lived gluino is
the existance of neutral hadrons containing them.  Generically,
hadrons containing a single gluino are called $R$-hadrons\cite{f:23}.
The lightest of these would be the neutral, flavor singlet $g \tilde{g}$
``glueballino'', called $R^0$.  There would also be $R$-mesons,
$\bar{q}q \tilde{g}$, and $R$-baryons,$qqq \tilde{g}$, with the $\bar{q}q$
or $qqq$ in a color octet.  Unlike ordinary baryons which are unable
on account of fermi statistics to be in a flavor singlet state, there
is a neutral flavor-singlet $R$-baryon, $uds \tilde{g}$, called $S^0$
below.  It should be particularly strongly bound by QCD hyperfine
interactions, and probably is the lightest of the
$R$-baryons\cite{f:51,f:52}, even lighter than the $R$-nucleons.

The strategy pursued here is to identify production and detection
mechanisms for the $R^0$ for which {\it reliable} rate estimates can be
made, so that searches which are sufficiently sensitive will
definitively rule them out or find them.  First, we use theoretical
arguments to estimate $R$-hadron masses as a function of gluino mass.
Then experiments are proposed to settle the question.

\section{$R$-hadron mass estimates}
\label{mass}
\hspace*{2em}

If the gluino is heavier than $\sim 3.5$ GeV, then the $R^0$ and $S^0$
will have masses approximately equal to the mass of the gluino.  For
the window $ m_{\tilde{g}} \,\raisebox{-0.13cm}{$\stackrel{\textstyle<}
{\textstyle\sim}$}\, 1.6$ GeV, lattice gauge theory (lgt)
should be used to determine the hadron spectrum, and hopefully the
necessary calculations will be done soon.  However we can get a rough
idea without it, as follows.  Let us begin by estimating hadron masses
if the gluino is as light as possible.

If the gluino were massless, the spectrum would be expected to contain
an unacceptably light\cite{ev,sv} flavor-singlet goldstone boson
associated with the spontaneous breaking of the non-anomalous linear
combination of quark and gluino chiral U(1) symmetries.  For three light
flavors of quarks the non-anomalous axial current is\footnote{The fields
appearing in this expression are left-handed Weyl spinors and a sum over
indices is understood.  $i$ labels the three light quark flavors and $j$
and $a$ label the 3 quark and 8 gluino color degrees of freedom.}:
\begin{equation}
J^5_{\mu} = \frac{1}{\sqrt{26}} \left\{\bar{q}^{i,j}_L \gamma_{\mu}
q^{i,j}_L - \bar{q^c}^{i,j}_L \gamma_{\mu} {q^c}^{i,j}_L -
\bar{\lambda^a} \gamma_{\mu} \lambda^a \right\}.  \label{j5nonanom}
\end{equation}
We can obtain a theoretical lower bound on the gluino mass by
identifying the $\eta'$ with this pseudogoldstone boson\footnote{This
possibility was suggested in ref. \cite{f:51} but not developed in
quantitative detail as is done here.}.  The flavor singlet pseudoscalar
which gets its mass from the anomaly would then be identified with a
more massive state, which will be discussed below.

If this were the correct description of the $\eta'$, its quark
content would be reduced by a factor of $\frac{18}{26} \approx 0.7$ in
comparison to the usual picture.  Interestingly, this seems not to be
ruled out by existing constraints.  Sound predictions for the $\eta'$,
avoiding model dependent assumptions such as the relation between
$F_1$ and $F_8$, are for ratios of branching fractions to final states
which couple to the quark component\cite{chanowitz:etaprime}.  These
ratios are insensitive to the presence of a gluino or gluonic
component.  Absolute predictions are highly sensitive to theoretically
incalculable hadronic effects, due to the very restricted phase space
for the $\eta'$ to decay through strong interactions.  This means that
rates which could potentially determine whether the $\eta'$ has a
$30\%$ gluino component, in practice cannot be predicted reliably
enough to be useful.\footnote{A possible way to discriminate is to
study the production of the various pseudoscalars in $J/\Psi$ decay.
G. Farrar and G. Gabadadze, in preparation.}

Assuming tat the $\eta'$ is the pseudogoldstone boson connected to
the spontaneous breaking of the conserved axial current
(\ref{j5nonanom}), like the $K^+$ for $J^5_K = \frac{1}{\sqrt{6}}
\left\{\bar{u}^j_L \gamma_{\mu} s^j_L - \bar{u^c}^j_L \gamma_{\mu}
{s^c}^j_L \right\}$, standard current algebra manipulations lead to
predictions for $m_{\eta'}^2 f_{\eta'}^2$ and $m_{K}^2 f_{K}^2$.
Taking their ratio, neglecting $m_u$ and $m_d$ in comparison to $m_s$,
and solving for $m_{\tilde{g}}$ leads to:
\begin{equation}
m_{\tilde{g}} \approx
\frac{m_s}{4}\frac{<\bar{s} s>}{<\bar{\lambda} \lambda>}\left[13
\left( \frac{m_{\eta'}^2 f_{\eta'}^2}{m_{K}^2 f_{K}^2 }\right) - 3
\right].
\label{mgluino}
\end{equation}
With $f_{\eta'} \approx f_K$ this gives
$m_{\tilde{g}} \sim 11 \frac{<\bar{s} s> }{<\bar{\lambda} \lambda>}
m_s$.  Since the QCD attractive force between color octets is greater
than that between triplet and antitriplet, $<\bar{\lambda} \lambda>$
is presumably larger than $<\bar{s}s>$.  Most-attractive-channel
arguments\cite{drs} suggest that the condensates depend exponentially
on the Casimirs of the condensing fermions so that since $C_8/C_3 =
9/4$, $<\bar{\lambda} \lambda>$ could be an order of magnitude or more
larger than $<\bar{s}s>$.  Thus pending lattice calculations of
$<\bar{\lambda} \lambda>$ or $m(\eta')$ as a function of gluino mass
and without gluinos, the phenomenological analysis should be general
enough to include a gluino as light as $\sim 100$ MeV or less.  In
this case the $R$-hadron properties are about the same as they would
be for a massless gluino.

If the gluino were massless, the mass of the $R^0$ should be $1440 \pm
375$ MeV, i.e., about $1 \frac{1}{2}$ GeV, as follows.  Consider
supersymmetric SU(3) Yang Mills theory.  Since supersymmetry in this
theory does not break dynamically\cite{witten}, hadrons must fall into
degenerate supermultiplets.  The massive chiral supermultiplet
containing the $0^{++}$ glueball also contains a $0^{-+}$ (the lowest
$\tilde{g} \tilde{g}$ bound state) and two spin-$\frac{1}{2}$ states,
namely the two helicities of the $R^0$ (the $g \tilde{g}$ bound state).
At the classical level this theory has a chiral U(1) phase
invariance since the gluinos are massless, like the chiral U(1) of
ordinary QCD with massless quarks.  This symmetry is clearly not realized in
the hadron spectrum, since the $R^0$ is degenerate with the massive
glueball.  Nor is there a goldstone boson associated with the breaking
of this U(1) symmetry since the pseudoscalar $\tilde{g} \tilde{g}$ bound
state is also degenerate with the glueball.  This is not paradoxial
for the same reason that in ordinary QCD we can accomodate the $\eta'$
mass.  Namely, the axial U(1) current has an anomaly so that
non-perturbative effects give the pseudoscalar $\tilde{g} \tilde{g}$
bound state a mass.  The chiral U(1) symmetry is explicitly broken by
quantum effects so that Goldstone's theorem is
circumvented\footnote{It is interesting that supersymmetry relates the
mass produced by non-perturbative effects through the anomaly to the
mass-gap for the glueball coming from confinement, suggesting that
confinement is essential to the understanding of the mass of the
$\eta'$ even in ordinary non-susy QCD.}.

Now consider hadron masses in QCD with three light quarks and a
massless gluino.  The mass of the $0^{++}$ glueball is
predicted\cite{weingarten:glueballs} using lattice QCD in quenched
approximation (i.e., QCD with only gluons and no quarks or gluinos) to
be $1440\pm 110$ MeV.  In ordinary lattice QCD, with three light
quarks but no gluinos, the quenched approximation is commonly taken to
be valid at the $10-15 \%$ level for hadron masses\footnote{See
ref. \cite{quenched} for a critical discussion of quenched
approximation.}.  Since the 1-loop beta function for QCD with no light
quarks but an octet of gluinos is the same as for QCD with three light
quarks, one can expect that the error for quenched approximation in
the supersymmetric Yang Mills theory is also $10-15 \%$, and that the
quenching error with both quarks and gluinos is $\sim 15-25 \%$.  If
this is so, then a full lattice calculation in QCD with 3 light quarks
and a massless octet of gluinos would give a mass for the $R^0$ of
$\sim 1440 \pm 375$ MeV, where the lattice error of
ref. \cite{weingarten:glueballs} on the glueball mass was combined in
quadrature with the estimated error from the quenched approximation,
taken to be $25\%$ of $1440$ MeV to be conservative.  As
we shall see below, it is much more difficult to detect an $R^0$ with
mass $\sim 1065$ GeV than one with mass $\sim 1800$ GeV.  Thus a lgt
calculation which reduced the range of uncertainty on the mass of the
$R^0$ would be very helpful, especially if it showed we could ignore the
region close to 1 GeV.

In QCD extended by gluinos, the flavor singlet pseudoscaler which gets
mass from the anomaly is orthogonal to the anomaly-free current
(\ref{j5nonanom}), thus it is $70\%$ $\tilde{g} \tilde{g}$ and $30\% ~ u
\bar{u} + d \bar{d} + s \bar{s}$.  In the supersymmetric Yang Mills
theory discussed above, the pseudoscalar $\tilde{g} \tilde{g}$ state which
gets mass from the anomaly and is degenerate with the $0^{++}$ glueball
would have a mass of $1440 \pm 240$, adding the error in ref.
\cite{weingarten:glueballs} in quadrature with a $15\%$ error for
unquenching.  There is evidence for an ``extra'' flavor singlet
pseudoscalar present in the meson spectrum in the 1410-1490
region\cite{PDG,mark3,dm2}, which has a large coupling to
gluons\cite{f:93}.  If confirmed, it is an excellent candidate to be
the pseudoscalar whose mass comes from the anomaly, in the very light
gluino scenario.

To recapitulate, we have seen above that from purely theoretical
considerations we can at present only rule out $R^0$ and $S^0$ masses
below about 1100 MeV.  Having the lightest possible masses
requires both the $\eta'$ and extra pseudoscalar meson in the
1410-1490 MeV region to have large gluino components, but increasing
the gluino mass to $\sim 700$ MeV allows one to return to the
conventional phenomenology for the $\eta'$ and interpret the extra
state as a simple $\eta_{\tilde{g}}$, the lowest-lying $\tilde{g}
\tilde{g}$ bound state.  If gluinos are much heavier than this, one
needs another explanation for the extra state in the 1410-1490
region.

\section{Existing Experimental Limits}
\label{exptlims}
\hspace*{2em}

{}From the CUSB experiment, we infer\footnote{See footnote on
the first page of the present paper.} that the $\eta_{\tilde{g}}$ does not
lie in the 3-7 GeV range, so that the gluino would not be in the $\sim
1.5-3.5$ GeV range.  In order to compare to limits from other
experiments searching for $R^0$'s, we shall convert this limit to an
effective gluino mass using the relation
\begin{equation}
m(R^0) = 0.72 (1 +
e^{-\frac{m_{\tilde{g}}}{2}}) + m_{\tilde{g}} (1 - e^{-m_{\tilde{g}}}),
\label{mRmg}
\end{equation}
with all
masses in GeV. This is actually just a convention for making the
figure, but is physically reasonable in that it yields the
$m_{\tilde{g}}=0$ result of the previous section and in analogy with
mesons made of one light and one heavy quark associates an additive
confinement energy of about half the mass of a light-quark-meson
(here, of the $0^{++}$ glueball whose mass is $\sim 1.44 $ GeV) to the
light constituent (here, the gluon) of a light-heavy composite.

In another quarkonium decay experiment, the ARGUS group\cite{argus}
looked for events in which $\Upsilon' \rightarrow \gamma +
\chi_b(1^3P_1)$, followed by $\chi_b(1^3P_1) \rightarrow g \tilde{g}
\tilde{g}$, with one of the final $R$-hadrons decaying in a distance of
1-60 cm from the $e^+ e^-$ interaction point.  From the absence of
such events at a level predicted by pQCD they concluded that gluinos
in the mass range 1-4.5 GeV do not exist in the lifetime range to
which they were sensitive.  However perturbative QCD overestimates the
branching fraction $\chi_b(1^3P_1) \rightarrow g \tilde{g} \tilde{g}$ for
very light gluinos, since it fails to include
the effect of the substantial reduction in phase space arising from
the minimum invariant mass of a pair of $R^0$'s being about 3 GeV,
even when the gluino is massless (see section \ref{mass}).  To
determine whether the experimental sensitivity extends to a gluino
mass as low as 1 GeV as stated in ref. \cite{argus}, the
experiment should be reanalyzed using a more realistic model of
the branching fraction for $\chi_b(1^3P_1) \rightarrow g \tilde{g}
\tilde{g}$ in the non-perturbative portion of phase space.  The ARGUS
results, taken from Fig. 4a of ref. \cite{argus}, are plotted on the
figure using the above function to convert from their quoted gluino
masses to a common $R^0$ mass.  For the largest masses no conversion
is used, in order not to make the non-sensical claim that they can
exclude $R^0$'s which cannot be kinematically produced.

The best constraints beyond CUSB and ARGUS for long-lived gluinos in the
radiatively-generated range of up to $O(30)$ GeV come from searches
for new neutral particles.  Gustafson et al.\cite{gustafson} searched
for new hadrons with lifetimes greater than $10^{-7}$ sec, using
time-of-flight in a 590m long neutral beam at FNAL.  On account of
timing and energy resolution limitations, they were capable of
distinguishing a particle from a neutron only if its mass was
greater than 2 GeV.  From the limits of Gustafson et al, Dawson, Eichten
and Quigg\cite{deq} (DEQ) concluded that gluino masses in the 2-4 GeV
range could be excluded.  This experiment is therefore consistent with
CUSB and Bernstein et al (see below), and for $\tau_{\tilde{g}} > 10^{-7}$ sec
extends the lower end of the excluded mass range to 2 GeV as shown in
the figure.

The experiment of Bernstein et al.\cite{bernstein} places an upper
bound on the production cross-section of a neutral hadron produced in
400 GeV proton collisions, with mass in the range $1.5-7.5$ GeV, which
decays with a lifetime $(10^{-8} - 2 \times 10^{-6})$ sec to a 2- or
3-body final state containing a charged hadron.  They find $E \frac{d
\sigma}{d^3p}|_{90^o} \,\raisebox{-0.13cm}{$\stackrel{\textstyle<}
{\textstyle\sim}$}\, 5 \times10^{-35} \frac{\rm
cm^2}{\rm(GeV^2/c^3)}$ for mass of 1.5 GeV, and
$\,\raisebox{-0.13cm}{$\stackrel{\textstyle<}
{\textstyle\sim}$}\, 3 \times 10^{-32}
\frac{\rm cm^2}{\rm(GeV^2/c^3)}$ for 7.5 GeV, taking the most
sensitive lifetime value of $3 \times 10^{-8}$ sec.  Typical decays
would be $R^0 \rightarrow {\rm lsp} + \pi$('s) and $S^0 \rightarrow
{\rm lsp} + \Lambda^0 + \pi$('s) or $S^0 \rightarrow {\rm lsp} + N + K +
\pi$('s).  Since the $S^0$ has baryon number $+1$, it would be
expected to be produced mainly in the forward direction rather than at
$90^o$ where the experiment was done, so is not directly constrained
by this experiment.  However this experiment does constrain the possibility of
$R^0$'s.  For the light end of the mass range, a reasonably good
analog process which should be even more OZI-suppressed, is $p p
\rightarrow \bar{p} X$ whose invariant cross section is $\sim 10^{-27}
\frac{\rm cm^2}{\rm(GeV^2/c^3)}$\cite{bourquin_gaillard}, for similar
kinematics.  For a gluino mass of 3.5 GeV or larger, it is legitimate to
use perturbative QCD (pQCD) to compute the expected rate, as a
function of gluino mass.  This was done in tree approximation by
DEQ\cite{deq} for $m_{\tilde{g}} = 3$ GeV.  They predicted an invariant
cross section of $\sim 10^{-28} \frac{\rm cm^2}{\rm(GeV^2/c^3)}$ for
$p_{\perp} = 0$.  We can very crudely estimate the cross section for
production of a gluino of higher mass but $p_{\perp}=0$ by noting that
the cross section is mainly dependent on the combination $ m^2 +
p_{\perp}^2$.  The DEQ prediction for $m=3$ GeV and $p_{\perp} = 4$
GeV is $\sim 3 \times 10^{-34} \frac{\rm
cm^2}{\rm(GeV^2/c^3)}$ (see Fig. 44), which is the same as the
Bernstein et al limit for $m=5$ GeV and $p_{\perp} = 0$.  Thus
the Bernstein et al limit very roughly rules out $R^0$'s with mass
less than 5 GeV.  The range of lifetime sensitivity corresponding to
the cross section limits of Fig. 4a is shown in Fig. 4b, for $m=3$
GeV where it is $\sim 2 \times 10^{-8} - 2 \times 10^{-7}$ sec.  Since for a
fixed production rate the detector sensitivity depends mainly on
$\gamma \beta \tau$, and $\gamma \sim m^{-1}$, the range of maximal
sensitivity will shift upward, roughly in proportion to $m$, for $m>3$ GeV.
The range excluded by Bernstein et al is shown in the figure.  It is the
upper elongated region ending at 5 GeV.

The limits could in principle be somewhat tightened if there are charged
$R$-hadrons which decay only weakly to the $R^0$ or $S^0$, e.g., $R_{K^+}
\rightarrow R^0 + \pi^+$.  This will be the case if the mass
gap between charged $R$-pions and $R$-kaons and the $R^0$, and between
charged $R$-baryons and the $S^0$, is greater than the mass of the
corresponding kaon or pion.  Lattice calculations of the $R$-hadron
mass splittings as a function of gluino mass are badly needed here.
Bag model predictions for $R$-hadrons cannot be trusted since
parameters fixed to fit the ordinary hadrons may not be applicable to
$R$-hadrons, and furthermore bag model estimates have not been been
reliable for the glueball spectrum.  Nonetheless old bag model
estimates\cite{chanowitz_sharpe,f:51,f:52} suggest that for some
parameters there may not be enough phase space for $R_K \rightarrow K
+ R^0$ or $R_N \rightarrow K + S^0$.  Thus a search for charged
$R$-hadrons is worthwhile even though a null result would not exclude
gluinos.   Note that there is no relation between the lifetimes of the
$R^0$ and $S^0$ and lifetimes of charged $R$-hadrons, since the latter
decay to the $R^0$ and $S^0$ through conventional weak interactions
and would be expected to have a lifetime comparable to weakly decaying
hadrons of a similar mass, i.e., $10^{-10} - 10^{-13}$ sec for masses
in the range 1 - 5 GeV.  Briefly, the experimental constraints would be:

\begin{itemize}
\item  Cutts et al\cite{cutts} use time of flight to exclude lifetimes
greater than $\sim(2-5) \times 10^{-8}$ sec, for charged particles
with masses in the 4-10 GeV range.
\item  Bourquin et al\cite{bourquin} search for decaying particles in
the CERN hyperon beam, extending the excluded range for new charged
particles to cover the 2-4 GeV mass range, for lifetimes of order
$10^{-9}-10^{-8}$ sec.
\item  Charged $R-$hadrons having mass of
the same order of magnitude as the $D$ or $B$ mesons must have a lifetime too
short or long to decay in vertex detectors used to measure $D$ and $B$
lifetimes.
\item  There is a CDF limit on the existance of
charged hadrons having $\gamma \tau
\,\raisebox{-0.13cm}{$\stackrel{\textstyle>}
{\textstyle\sim}$}\, 10^{-7}$
sec\cite{cdf:stablechlim}, but it only addresses masses greater than
50 GeV because the present detector has time resolution at the
nanosecond level.

\end{itemize}
Otherwise the constraints on charged $R$-hadrons are poor and the
coverage is surprisingly spotty.  It must be reemphasized, however, that
even if strong-interaction-stable $R$-hadrons exit, one cannot immediately
apply these experimental constraints on the allowed regions for their
mass and lifetime to the limits in the figure, because there is no
direct relation between the lifetime of a charged $R$-hadron and that
of the $R^0$.

If the gluino lifetime is long because the squark mass is much larger
than $m_W$, then beam dump
experiments\cite{f:23,charm,ball,bebc,akesson}, which look for the
reinteraction of the lsp in a neutrino detector, become ineffective
because the lsp cross section falls as $\left( \frac{m_W}{M_{sq}}
\right)^4$.  Even if the lsp cross section is not too small, the
gluino must decay before losing energy in the dump, e.g., in 10 cm in
the Ball et al FNAL beam dump experiment\cite{ball,f:55}, i.e.,
requiring a lifetime $ \,\raisebox{-0.13cm}{$\stackrel{\textstyle<}
{\textstyle\sim}$}\, 5 \frac{m_{\tilde{g}}}{1 {\rm GeV}} 10^{-11}$ sec.
Likewise, the BEBC experiment\cite{bebc} observes that if
$\tau_{\tilde{g}} \,\raisebox{-0.13cm}{$\stackrel{\textstyle>}
{\textstyle\sim}$}\, 5 \times 10^{-11}$ sec the gluino decay does not
occur before interaction, ``severely degrading the photino flux
reaching our detector.''  For massless photino they model this effect,
but in general beam dump experiments need to be analyzed in terms of
the three parameters $m_{\tilde{g}},~\sigma_{\rm lsp}$ and
$\tau_{\tilde{g}}$.  Beam dump experiments cannot be used to exclude
regions of the gluino mass-lifetime plane without further assumptions
which are not in general appropriate to our case, except for gluinos
with lifetimes shorter than about $ 5 \times 10^{-11}$ sec.  The
HELIOS experiment\cite{akesson} explicitly addresses direct production
of WINP's, and not long-lived gluinos, since it requires that no
energy degradation occur in the dump.

The possibility of large gluino mass is at present only addressed by
collider missing energy searches that detect the existance of a gluino which
decays inside the apparatus with a substantial portion of its energy
going to the lsp which is very weakly interacting and escapes.
Indeed, this is the classic gluino signal\cite{f:24}.  The CDF missing
energy search\cite{cdf:gluinolim2} is sensitive to gluinos which decay
within about 1 meter of their production, i.e., having $\frac
{E_{\tilde{g}}}{m_{\tilde{g}}} \tau_{\tilde{g}}
\,\raisebox{-0.13cm}{$\stackrel{\textstyle<}
{\textstyle\sim}$}\, 3 \times 10^{-9}$ sec.
They require the missing transverse energy to be greater than 40 GeV.
To get a very rough idea of their regime of sensitivity (which could
be determined more accurately by modeling the energy spectrum of the
produced gluinos) we can take as a typical event the case in which the
gluino is emitted at $45^o$ and assume it decays giving 1/3 its energy
and momentum to the lsp which escapes with the minimal transverse
energy to satisfy their cuts.  In this case, the actual energy of the
decaying gluino would be $3 \cdot \sqrt{2} \cdot 40 = 170 $ GeV,
ignoring gluino, quark and lsp masses.  Thus gluinos with lifetimes
longer than about $2 \times 10^{-11} \left(\frac{m_{\tilde{g}}}{1 {\rm
GeV}}\right) $ sec, would not be efficiently detected in the CDF
search.  They do not investigate masses lower than 20 GeV, where they
lose efficiency on account of the acoplanarity and missing $E_T$ cuts.

The UA1 missing energy search\cite{ua1} claims to be sensitive enough
to exclude masses as low as 4 GeV.  Although a gluino lifetime is not
included in their efficiency monte carlo, they state that they believe
they are fully sensitive to gluinos whose lifetime is shorter than
$10^{-10}$ sec.  This agrees with the crude estimate given above for
the CDF experiment, for $m_{\tilde{g}} = 5$ GeV.  Nonetheless,
especially for the high mass end of the UA1 experiment, a monte carlo
is needed to know the lifetime sensitivity as a function of mass.  For
simplicity, and to be conservative, we will use the estimate
$2 \times 10^{-11} \left(\frac{m_{\tilde{g}}}{1 {\rm GeV}}\right) $ sec
for both CDF and UA1.

To summarize this section, gluinos in the mass range $\sim 1.5-3.5$
GeV are absolutely excluded (CUSB).  Lighter gluinos are allowed, as
long as the $R^0$ lifetime is not in the range $2 \times 10^{-6}-
10^{-8}$ sec if the $R^0$ mass is greater than 1.5 GeV (Bernstein et
al), or the range $> 10^{-7}$ sec if its mass is greater than 2 GeV
(Gustafson et al).  Gluinos with mass around 4 GeV or above, must have
a lifetime longer than about $\sim 2 \times 10^{-11} \left(
\frac{m_{\tilde{g}}}{1 {\rm GeV}}\right) $ sec (UA1,CDF), with the
ranges $> 10^{-7}$ sec (Gustafson), $2 \times 10^{-6}-10^{-8}$ sec
(Bernstein et al) and $\sim 10^{-10}$ sec (ARGUS) ruled out for masses
in the vicinity of 4-5 GeV.  The figure is an attempt to summarize
these results, combining experiments which report results directly in
terms of $m(R^0)$ with those characterized by limits on $m_{\tilde{g}}$
by use of eqn.  (\ref{mRmg}).  Given the primitive nature of
eqn. (\ref{mRmg}) and the $\pm 375$ MeV uncertainty on the $R^0$ mass
when the gluino is massless (section \ref{mass}), as well as the very
rough methods used to extract the ranges of mass and lifetime
sensitivity for the various experiments, a
$\,\raisebox{-0.13cm}{$\stackrel{\textstyle>}
{\textstyle\sim}$}\, 20 \%$ uncertainty
should be attached to all the boundaries shown in this figure.

\section{Theoretical Comments: Gluino Lifetime and Production
Estimates}
\label{lifetime}
\hspace*{2em}

How natural is it from a theoretical point of view for an $R^0$ in the
mass range $1.5 - 2.5$  GeV to have a lifetime longer than $2 \times
10^{-6}$ sec, or for an $R^0$ with mass
$\,\raisebox{-0.13cm}{$\stackrel{\textstyle>}
{\textstyle\sim}$}\, 5$ GeV to have a lifetime longer
than $\sim 2 \times 10^{-11} \left( \frac{m_{\tilde{g}}}{1 {\rm
GeV}}\right) $ sec?  For the higher mass range the $R^0$ and gluino lifetimes
can be taken to be approximately the same, since for a relatively
massive state one can ignore the effects of confinement on the overlap
of the initial and final states, and the modifications to phase space
from the hadron masses.  For the low end of the range, if the lsp
mass is low compared to the gluino mass, one could either
argue by analogy to known hadron decays\cite{f:23} or, following
Franco\cite{franco}, take the $R^0$ lifetime to be that of a gluino of
$\sim \frac{3}{4}$ of its mass.  For the interesting case that the lsp
mass is a significant fraction of $m(R^0)$, tools have not yet been
developed which allow us to reliably estimate the resultant
suppression in the decay rate.

The decay rate for an unconfined gluino to decay to the lsp and a $
\bar{q} q $ pair can be obtained as follows.  In general, the lsp is a
superposition of the fermionic partners of the neutral SU(3) and U(1)
gauge (w3-ino and bino) and Higgs bosons (higgsinos).  However it is
shown in ref. \cite{f:96} that when gaugino masses are all radiatively
generated the higgsino component of the lsp is in fact less than $1\%$
in amplitude.  Thus we can approximate the lsp wavefunction as $
cos\theta |\tilde{b}> + sin\theta |\tilde{w}_3>$.  The decay rate of the
gluino assuming the lsp to be a photino was given in
ref. \cite{hk:taugluino}, so we need only replace $e_q^2$ appearing in
their expression by $(\frac{sin \theta}{sin \theta_W})[I_z + z
\frac{Y}{2}]^2$, where $z=\frac{tan \theta_W}{tan\theta}$, and average
over left and right handed contributions.  Thus the total rate for
gluino to decay to the lsp and a $u \bar{u},~d \bar{d}$, or $s
\bar{s}$ pair, ignoring the quark masses is:
\begin{eqnarray}
& \Gamma_{\tilde{g}} = \frac{\alpha_s \alpha_{em} (1 - \frac{2}{9}z +
z^2) m_{\tilde{g}}^5}{128 \pi M_{sq}^4} (\frac{sin \theta}{sin
\theta_W})^2 \times & \\
& [(1-y^2)(1 + 2y - 7y^2 + 20y^3 - 7 y^4 + 2 y^5 + y^6) + 24y^3(1 - y
+ y^2)log(y)], & \nonumber
\label{taugluino}
\end{eqnarray}
where $y=\frac{m_{\rm lsp}}{m_{\tilde{g}}}$.  We have taken $M_L^{sq} =
M_R^{sq} = M_{sq}$ for simplicity.  The $\theta$ dependent
factor ranges from 1, for a light neutralino in the low-$\mu$
region where $\theta \approx \theta_W$, to $(cos \theta_W)^{-2}$ for
a heavy neutralino in the high-$\mu$ region where $cos \theta
\approx 1$.  Thus for a rough estimate we take this factor to be 1.
We also take $\alpha_s \sim 0.1$ and $\alpha_{em} = 1/128$, since the
relevant scale is the squark mass.  Then, for instance with a massless
lsp, the squark mass must be greater than $\sim 2$ TeV for a gluino
with effective mass in the 1-1.5 GeV mass range to have $\tau_{\tilde{g}}
\geq 2~10^{-6}$ sec.  If instead the lsp mass is 90\% of the gluino
effective mass, the squark mass must only be greater than about 200 GeV.
For a gluino of mass 5 GeV, the UA1 bound is most relevant.  For lsp
mass of zero or $0.9 m_{\tilde{g}}$ one finds that the squark mass must be
greater than 1 TeV or $\sim 130$ GeV, respectively.  These squark masses
increase to 6 TeV or 670 GeV for a 15 GeV gluino.  As shown in ref.
\cite{f:96}, when gaugino masses arise radiatively, these conditions are
naturally accomodated in much of parameter space.

It is also worth noting that absolute stability is a real possibility
for the $S^0$, since the mass difference between it and the lsp must be
greater than the sum of proton and electron masses for it to decay.  If
it binds to nuclei, this would be ruled out experimentally by the
sensitive searches for exotic isotopes, at least for some mass
regions\cite{muller}.  However one would expect a repulsive, not
attractive, interaction between a nucleus and the flavor-singlet $R^0$
or $S^0$ , since the intermediate state created when they exchange
mesons with a nucleon has a much higher energy\footnote{Unlike the
binding of a nucleus where exchange of mesons between pairs of nucleons,
each of which can absorb or emit an $I=1$ meson and remain a nucleon,
leads to intermediate states close in energy to the original state.}.

Anomalous signals in extensive air showers and underground
muons seemingly coming from Cygnus X-3 are consistent with the
intermediate particle being a neutron, except that the neutron decays
too quickly to make the long trip\footnote{See, e.g., ref. \cite{bei}
for a summary.}.  Long-lived $R^0$'s were investigated\cite{bi}, but
discarded\cite{ov} on account of the mistaken belief that they would
imply a long lived charged $R$-proton which is ruled out by, e.g.,
ref. \cite{muller}.  If the present quiet of Cygnus X-3 is only a
cyclical phenomenon and such events are observed again in the future,
an $S^0$ interpretation should be seriously considered.

Turning now to cross section calculations, I am not aware of any recent
pQCD calculations of gluino production at a hadron collider, except for
very massive gluinos.  The old analyses\cite{deq,bhk} should be updated,
making an attempt to estimate the uncertainty in the gluino
distribution, as well as including 1-loop corrections which have proved
very important for ordinary pQCD predictions.  From deep inelastic and
Drell-Yan experiments, the quark and antiquark
distributions are reasonably well fixed.  Direct photon production
gives information on the gluon distribution function, so the momentum
sum rule then provides some constraint on the gluino distribution.
The naive argument\cite{f:9} which leads to behavior $(1-x)^7$ for the
sea-quark distribution functions at large-$x$, leads to the same
behavior for the gluino distribution function.  Since the $R^0,
{}~\eta'$, and $\eta_{\tilde{g}}$ masses are so much larger than pion masses,
one would expect that the low-$Q^2$ gluino distribution functions are
smaller than those of the sea-quarks.  However since the 1-loop beta-function
for gluinos is the same as for 3 flavors of light quarks, the gluino
distribution function evolves as rapidly as all three quarks together,
so a light gluino would become an important component of the
nucleon at larger $Q^2$.

Although the gluon and gluino distribution functions are individually
difficult to determine well, without assumptions as to their
functional form for the entire $x$ range, their sum is much better
determined\footnote{Comparably to the determination of the gluon
distribution function, when the gluino possibility is ignored.}.
Since both gluons and gluinos give rise to gluino jets, the actual
prediction for $R^0$ production is relatively stable.  If the existance
of gluinos were established, the ratio of events with 1 and 2 $R^0$'s
would allow the ratio of gluino and gluon distributions to be constrained.
Demanding consistency of pQCD predictions with observed jet production
may also allow the gluino distribution function to be further
constrained, since the amplitudes for gluinos to produce jets differs
from those for quarks or gluons to produce jets.

\section{Indirect Evidence Regarding Light Gluinos}
\label{indirect}
\hspace*{2em}

For years it has been recognized that in principle the running of
$\alpha_s$ is sensitive to the presence of gluinos\footnote{For the
first discussion of this, see ref. \cite{f:32}}.
In deep inelastic scattering experiments the ambiguity introduced by
higher twist contributions is too large to allow one to decide between
QCD with and without gluinos\footnote{Thus comparison of the
values of $\alpha_s$ from deep-inelastic scattering and $Z^0$ decay
are inconclusive, although suggestive\cite{jk}.}.
Gluinos modify the $e^+ e^-$ annihilation cross sections only in order
$\alpha_s^2$, by providing an additional source of 4-jet events and
making virtual corrections to 2-jet events.  The possibility of infering
or excluding gluinos directly from LEP event characteristics was
discussed in ref. \cite{f:82}, where the sensitivity to the
as-yet-uncalculated 1-loop corrections was shown to be too great to
allow one to decide between ordinary QCD and QCD with massless
gluinos\footnote{More recent articles on this subject have come to the
same conclusion\cite{m-ts,opal4jet}.}.

The reason that it is generally difficult to discriminate between QCD
with and without gluinos is because adding gluinos to the theory
modifies it in competing ways which tend to cancel.  For instance the
value of $\alpha_s$ at LEP is obtained by fitting QCD predictions for
various aspects of event shapes and extracting the value of $\alpha_s$
which gives the best fit.  Gluinos are an additional source of 4-jet
events, but at the same time $\alpha_s$ runs more slowly when there
are gluinos.  This means that, for a given value of $\alpha_s(M_Z)$,
the typical value of $\alpha_s(Q_{eff})$ in multi-jet events is lower
than it would be for QCD without gluinos, which tends to reduce the
number of multi-jet events\footnote{Ref.
\cite{enr} correctly emphasized the need to extract $\alpha_s$ from data
with and without gluinos before evaluating the consistency of the
running between different energy scales, with and without gluinos.
However that analysis only includes the virtual corrections to the
running of $\alpha_s$ in LEP events and not the effect of real gluino
jet production which is of the same order, so is incomplete.  Bryan
Webber and I (unpublished) tried to see if we could find some systematic
preference of the LEP data for QCD with and without gluinos by looking
at the entire menagerie of quantities from which $\alpha_s$ is
extracted.  We found that the only region in which there was a
significant difference in predictions with and without gluinos is
precisely the region in which hadronization is most important, and for
which the 1-loop corrections to the 4-jet cross section (which are not
yet available) are crucial.}.

Just as the effects of gluinos tend to cancel at LEP, one cannot
simply say that the number of jets predicted at the Tevatron will be
increased by such-and-such an amount, since if there are light gluinos
they will be present in the hadron structure functions and will use
some of the ``room'' for gluons, reducing the production of
conventional jets to some extent\footnote{See ref. \cite{f:82} for
a discussion of the difficulty of using hadron collider
jet cross-section
characteristics to infer or exclude the existance of long-lived gluinos
given the present level of theoretical precision.}.  To address the
possibility of light gluinos by their effects on jets or the running of
$\alpha_s$, one must a) compare predictions for actual experimental
observables with and without gluinos and not try to compare derived
quantities such as $\alpha_s$ and b) {\it fully} incorporate gluinos
into the analysis, including their effects on distribution functions.

Recently, there have been a number of attempts to make the kind of
careful analysis which would be necessary to obtain reliable indirect
information of the possibilty of light gluinos.  Ref. \cite{RtauRZ}
used theoretical predictions for the hadronic branching fractions
$R_{\tau}$ and $R_Z$ with and without gluinos, to extract
$\alpha_s(m_{\tau})$ and $\alpha_s(m_{Z})$ with and without gluinos,
then checked whether the running of $\alpha_s$ between these values
was consistent with what QCD predicts, with and without gluinos.  The
main difficulty with this approach is the issue of how to treat the
effect of a light or massless gluino on $\tau$ decay, and also the
question of the validity of neglecting non-perturbative contributions
of order $\frac{1}{m^2}$ as is done in ref. \cite{bnp}.  The latter
issue is discussed in ref. \cite{altarelli:hanoi}, where it is argued
that unless the validity of neglecting $\frac{1}{m^2}$ corrections is
established, the error should be taken to be twice that assigned in
the ``nominal'' case of ref. \cite{RtauRZ}, i.e., that $R_{TH} = 2$ is
appropriate for the ref. \cite{RtauRZ} analysis.  With respect to the
former issue, since the invariant mass of an $R^0$ pair and of the
$\eta_{\tilde{g}}$ is too large to contribute significantly to $\tau$ decay,
independent of the gluino mass, the gluino contribution should be
neglected when determining $\alpha_s(m_{\tau})$, as is done for the
charm quark.  This can be implemented\footnote{M. Schmelling and
R. St.Denis, private communication.} by using an ``effective'' gluino
mass $\,\raisebox{-0.13cm}{$\stackrel{\textstyle>}
{\textstyle\sim}$}\, m_{\tau}/2$ when using their fig. 2.  One then finds that
even at 90\% confidence level there is no excluded region of gluino
mass from this analysis when $R_{TH} = 2$.

For other attempts to study indirect evidence for light gluinos see,
e.g., refs. \cite{clavelli,clavetal}.

\section{Proposals for Experiments}
\hspace*{2em}

Now let us turn to the question of how to establish or rule out the
existance of new light hadrons, $R^0$ or $S^0$.  One method, proposed
years ago\cite{f:51}, is to look for exclusive reactions such as $K^-p
\rightarrow R^0 S^0$, followed by elastic scattering of the $R^0$ and
$S^0$ off protons.  With accurate measurements of the $R^0$ and $S^0$
production angles, and measurement of the recoil proton momenta in the
secondary $R^0 p \rightarrow R^0 p$ and $S^0 p \rightarrow S^0 p$
scatterings, there is in principle one more equation than unknowns and
the masses of the $R^0$ and $S^0$ can both be determined.  Using a
hydrogen bubble chamber would seem to work nicely for observing the
initial and secondary scatterings, but a high efficiency for
identifying $K^0_L$'s and neutrons would be desirable to reduce
background, so this may not be the optimal approach.  The interaction
lengths of the $R^0$ and $S^0$ are probably somewhat shorter than for ordinary
mesons and baryons, on account of the greater color charge of the
gluinos as compared to quarks and on account of the $S^0$ having 4
constituents rather than 3 for a normal baryon.  The candidate events
should show a threshold behavior consistent with the measured $R^0$
and $S^0$ masses, which would corroborate the validity of the overall
picture.  Note that this experiment is sensitive to gluinos with any
lifetime long enough that the $R^0$ and $S^0$ rescatter before
decaying, so that it is complementary to the experiment of Bernstein
et al. and sensitive to lower masses than Gustafson et al.  However
this method has two important weaknesses: First, the
cross section may be very small, since one is asking for a very exotic
final state to be produced in an exclusive mode.  Second, it is not
possible to reliably calculate the cross section so that one cannot
establish a level of sensitivity adequate to definitively exclude the
phenomenon.  Unfortunately it is also a demanding, single-purpose
experiment and theoretical prejudice has favored heavy gluinos, so
that experimenters have not looked just to see if something might be
there\footnote{Recently, Carlson and Sher\cite{cs} proposed searching for the
decays of gluinos following their photoproduction at CEBAF.  This is
an excellent experiment, since something may be found.  However it
does not satisfy the present criterion of being useful for excluding a
light gluino, since the relatively low invariant mass range which can
be probed at CEBAF means that the non-perturbative effects of $R^0$
and $\eta_{\tilde{g}}$ masses will suppress the signal and the
calculations of the production rates are therefore not sufficiently
reliable to allow exclusion.  They report results for the effective
gluino mass being taken to be 1 and 1.5 GeV, and the dramatic rate of
decrease with effective gluino mass reflects the sensitivity to this
effect.  To obtain reliable inclusive cross sections for production of
light particles from pQCD, one must impose a $p_{\perp}^{min}$ cut.
The event rates they quote are so large that this may be possible, but
as long as their signal is the decay of the gluino, the proposed
experiment can only be used to examine a limited lifetime region.}.

Here I propose other experiments which also do not rely on
observing the decay of the $R^0$ or $S^0$ and are thus able to rule
out or observe long-lived gluinos, but which do not have the
difficulties of the one discussed above.  Except in the forward direction,
we expect that $S^0$ production is much smaller than $R^0$ production,
so let us ignore $S^0$'s for simplicity.  By working at high energy,
exotics can be produced relatively easily and inclusive cross-sections
can be reliably computed from perturbative QCD in appropriate
kinematical regions.  The cross section for producing $R^0$'s is
essentially just the gluino jet cross section, since all gluino jets
end in an $R^0$ (or, rarely $S^0$) because other $R$-hadrons
eventually decay to these\footnote{Or {\it very} rarely, gluinos
from independent jets can annihilate, but at this order one must also
consider jet evolution which produces gluinos.}.  The gluino-jet
cross-section is approximately 10\%\cite{f:82} of the total jet cross
section, so that it is actually quite common for a Tevatron collider
or fixed target event to contain an $R^0$ pair.  $p_t$ cuts can be
imposed to insure that perturbative QCD event generators can be
reliably used to compute the expected rate, even for light gluinos.
Showing that there are no such events at a level of $4\sigma$ below
the prediction, would then convincingly rule out the existance of
these gluinos\footnote{Care must be taken to realistically estimate
the theoretical uncertainty, including that from the distribution
functions and neglect of higher order corrections to the partonic
scattering amplitudes, which in ordinary QCD have proven to be larger
than originally estimated.}.

Basically the idea is an outgrowth of the suggestion of ref.
\cite{f:51}, but sacrificing the additional constraints of exclusive
production in favor of the higher rate and reliable calculability of
high energy inclusive production.  A high energy beam from an
accelerator is incident on the primary target.  This produces a
neutral beam containing neutrons, kaons, hyperons, and possibly
$R^0$'s and $S^0$'s.  This beam illuminates a secondary target in
which an elastic scattering $R^0 p \rightarrow R^0 p$ may occur.  Measuring the
momentum of the recoil proton and the angle of the produced $R^0$ (by
observing {\it its} interaction, which need not be elastic)  gives
enough constraints to solve for $m_R$, if indeed the reaction is
elastic.  Knowing the visible energy of the final particles in the
secondary scattering of the produced $R^0$ can help choose between multiple
solutions and help discard events in which the primary scattering is
not elastic.\footnote{I am grateful to T. Devlin for making these points.}
Of course the background due to other reactions, especially $n~p
\rightarrow n~ p$ or $K^0_L p \rightarrow K^0_L p$, or inelastic
scattering, will be quite severe even after vetoing on extra charged
particles and $\pi^0$'s, so excellent resolution is crucial.

Timing could be used to measure $p/E$ of the incident neutral.  With
this information, one would have an over-constrained system of equations
without relying on the secondary scattering being elastic and one
could verify that the initial reaction was indeed $R^0 p
\rightarrow R^0 p$ as well as determine the $R^0$ mass.  If the
$R^0$ is sufficiently heavy, one can get adequate resolution with nsec
accuracy using the beam buckets without being forced to put the
secondary target so far away that the loss of solid angle would be
intolerable\footnote{Keeping the distance between the two targets as small as
possible is also desirable from the standpoint of being sensitive to
relatively shortlived $R^0$'s as well.}.  Modern $O(10)$ psec timing
could allow the lower mass regions to be investigated, except that it
requires tagging the initial $R^0$ production event, so entails a
reduction in rate.  Detailed monte carlo simulation is needed to
determine whether it is possible to cover the very-light
gluino regime, where the $R^0$ may be difficult to distinguish from a
neutron.  With many events, a discrepancy between the observed and
expected event characteristics such as angular distribution and rates
would be a useful diagnostic.  Another handle for some range of $R^0$
lifetimes would be a distance dependence of the anomalous events.

In the above discussion I focussed on the process $R^0 p \rightarrow R^0
p$ for identifying the $R^0$.  It is the most attractive option from a
theoretical point of view since its cross section is easiest to
estimate\footnote{The optical theorem relates the forward elastic cross
section to the total cross section.  Above the resonance region one
would expect $\sigma(R^0p)\sim \sigma (\pi p) \sim \sigma(pp)$ since the
confinement scale rather than the color charge of the valence
constituents, seems most important in determining the size of a system
of light, relativistic quarks or gluons or gluinos.  Using lattice gauge
theory, it might be possible to measure the color charge radius of the
$R^0$ or at least its ratio to that of the pion or nucleon, to improve
upon this crudest possible estimate.  Or one could use information from
lgt on glueball masses to try to constrain a bag model for color octet
constituents, and then determine their radius.}.  If one's goal is to try to
unambiguously exclude light gluinos, then one must use reactions which
can be estimated with some confidence both to produce and to detect
them.  However if one wants the most effective way to discover light
gluinos if they exist, one can consider other detection reactions such
as $R^0 p \rightarrow R_p \eta'$ or $R^0 p \rightarrow K^+ S^0$, whose
signature may be much more distinctive\footnote{I am indebted to W.
Willis for emphasizing this point.}.  In  the resonance region, such
cross sections can be very large.  Further work is needed to try to
estimate them.

A setup such as KTeV, where the distance between primary and secondary
targets (the regenerator) is 120m and the typical energy of the
long-lived neutrals is about 100 GeV, would be mainly sensitive to
lifetimes longer than $\sim 4~ 10^{-9}$ sec.  Thus if it can be used
for this purpose, it will be able to probe a large part of
the interesting lifetime range.

In a collider experiment, pair produced heavy gluinos would radiate
gluons and light quarks to produce jets containing ordinary hadrons
and an $R^0$.  For  sufficietly heavy $R^0$ and good timing
capabilities, one could in principle detect the time delay $p/E$ for
the late-arriving neutral particles to deposit energy in the
calorimeter.  Assuming each of them to be an $R^0$ which stopped in
the calorimeter, producing very light particles, the energy it
deposited in the calorimeter would be roughly of the same magnitude as
$p$ of the $R^0$.  Knowing $p$ and $p/E$, one could solve for $m_R$.
A detailed study of the conversion of an $R^0$'s momentum to the
energy deposition in the calorimeter (in particular the extent of the
fluctuations to be expected), is needed to see if this method is
feasible in practice.  Another way that the production of a pair of
heavy long-lived gluinos might be infered in principle, would be to
search for events in collider experiments in which fitting energy and
momentum conservation at the jet level requires two of the jets to be
given a large mass.

\section{Summary}
\hspace*{2em}

As is shown in ref. \cite{f:96}, if gaugino masses are generated by
loop effects, the gluino and lsp masses will be in the range from
$\sim 100$ MeV to $\,\raisebox{-0.13cm}{$\stackrel{\textstyle<}
{\textstyle\sim}$}\, 30$ GeV if the SUSY and ew symmetry breaking
scales are $\,\raisebox{-0.13cm}{$\stackrel{\textstyle<}
{\textstyle\sim}$}\, 10$ TeV.  Furthermore, in a substantial part of
parameter space the lsp is near in mass or heavier than the gluino, so
that long gluino lifetimes are natural.  The phenomenology of such
light, long-lived gluinos is the subject of the present paper.  Some
aspects of the phenomenology of the associated lsp are discussed in
ref. \cite{f:96}.  A very light gluino (mass of order a few hundred
MeV or less) is particularly attractive since it emerges naturally
when dimension-3 SUSY breaking operators are absent from the
low-energy theory, as is the case in hidden sector dynamical SUSY
breaking with no gauge singlets\cite{bkn}.  Consideration of the
pseudoscalar spectrum is shown to imply that the gluino mass
must be greater than $\sim 100$ MeV.  A very light gluino would
lead to new hadrons, the $R^0 ~ (g \tilde{g})$ and $S^0 ~ (uds\tilde{g})$,
with masses around $1 \frac{1}{2}$ GeV.  Experiments to definitively
rule out or discover them are possible but very challenging.  Existing
direct and indirect experimental constraints are reviewed and found
not to address the most interesting scenarios. Experiments directed at
the higher mass range are also mentioned.

\section{Acknowledgements}
I have benefitted from discussions with T. Banks, J. Bronzan, N. Christ,
J.  Conway, T. Devlin, H. Georgi, C. Lovelace, A. Masiero, A. Mueller,
M.  Schmelling, S. Somalwar, R. St.Denis and W. Willis.



\newpage


\begin{thebibliography}{10}

\bibitem{cdf:gluinolim2}
F.~Abe et~al.
\newblock {\em Phys. Rev. Lett.}, 69:3439, 1992.

\bibitem{f:51}
G.~R. Farrar.
\newblock {\em Phys. Rev. Lett.}, 53:1029--1033, 1984.

\bibitem{deq}
S.~Dawson et~al.
\newblock {\em Phys. Rev.}, D31:1581, 1985.

\bibitem{bgm}
R.~Barbieri, L.~Girardello, and A.~Masiero.
\newblock {\em Phys. Lett.}, B127:429, 1983.

\bibitem{f:96}
G.~R. Farrar and A.~Masiero.
\newblock Radiative gaugino masses.
\newblock Technical Report RU-94-38, Rutgers Univ., 1994.

\bibitem{keung_khare}
W.-Y. Keung and A.~Khare.
\newblock {\em Phys. Rev.}, D29:2657, 1984.

\bibitem{kuhn_ono}
J.~Kuhn and S.~Ono.
\newblock {\em Phys. Lett.}, 142B:436, 1984.

\bibitem{goldman_haber}
T.~Goldman and H.~Haber.
\newblock {\em Physica}, 15D:181, 1985.

\bibitem{tutsmunich}
P.~M. Tuts.
\newblock Limits on higgs boson and gluino masses from radiative upsilon
  decays.
\newblock In {\em Proceedings of the XXIV Int. Conf. on HEP, Munich}, 1988.
\newblock Contributed paper no. 723.

\bibitem{cusb}
P.~M. Tuts et~al.
\newblock {\em Phys. Lett.}, 186B:233, 1987.

\bibitem{f:93}
M.~Cakir and G.~R. Farrar.
\newblock Radiative decay of vector quarkonium: Constraints on glueballs and
  light gluinos.
\newblock {\em Phys. Rev.}, D50:3268, 1994.

\bibitem{f:23}
G.~R. Farrar and P.~Fayet.
\newblock {\em Phys. Lett.}, 76B:575--579, 1978.

\bibitem{f:52}
F.~Bucella, G.~R. Farrar, and A.~Pugliese.
\newblock {\em Phys. Lett.}, B153:311--314, 1985.

\bibitem{ev}
M.~I. Eides and M.~I. Vysotsky.
\newblock {\em Phys. Lett.}, 124B:83, 1983.

\bibitem{sv}
A.~V. Smilga and M.~I. Vysotsky.
\newblock {\em Phys. Lett.}, 125B:227, 1983.

\bibitem{chanowitz:etaprime}
M.~S. Chanowitz.
\newblock Technical Report LBL-29347, Massachusetts Institute of Technology,
  1990.

\bibitem{drs}
S.~Raby, S.~Dimopoulos, and L.~Susskind.
\newblock {\em Nucl. Phys.}, B169:373, 1980.

\bibitem{witten}
E.~Witten.
\newblock {\em Nucl. Phys.}, 202B:253, 1982.

\bibitem{weingarten:glueballs}
H.~Chen, J.~Sexton, A.~Vaccarino, and D.~Weingarten.
\newblock The scalar and tensor glueballs in the valence approximation.
\newblock Technical Report IBM-HET-94-1, IBM Watson Labs, 1994.

\bibitem{quenched}
M.~F.~L. Golterman.
\newblock Technical Report Wash. U. HEP/94-62, Washington University, 1994.

\bibitem{PDG}
Particle~Data Group.
\newblock {\em Phys. Rev.}, D45:S1, 1992.

\bibitem{mark3}
Z.Bai et~al.
\newblock {\em Phys. Rev. Lett.}, 65:2507, 1990.

\bibitem{dm2}
J.~E. Augustin et~al.
\newblock Technical Report 90-53, LAL, 1990.

\bibitem{argus}
ARGUS Collaboration.
\newblock {\em Phys. Lett.}, 167B:360, 1986.

\bibitem{gustafson}
H.~R. Gustafson et~al.
\newblock {\em Phys. Rev. Lett.}, 37:474, 1976.

\bibitem{bernstein}
R.~Bernstein et~al.
\newblock {\em Phys. Rev.}, D37:3103, 1988.

\bibitem{bourquin_gaillard}
M.~Bourquin and J.-M. Gaillard.
\newblock {\em Nucl. Phys.}, 114B:334, 1976.

\bibitem{chanowitz_sharpe}
M.~Chanowitz and S.~Sharpe.
\newblock {\em Phys. Lett.}, 126B:225, 1983.

\bibitem{cutts}
D.~Cutts et~al.
\newblock Search for long-lived heavy particles.
\newblock {\em Phys. Rev. Lett.}, 41:363, 1978.

\bibitem{bourquin}
M.~Bourquin et~al.
\newblock {\em Nucl. Phys.}, B153:13, 1979.

\bibitem{cdf:stablechlim}
F.~Abe et~al.
\newblock {\em Phys. Rev. Lett.}, 46:1889, 1992.

\bibitem{charm}
CHARM Collaboration.
\newblock Bounds on supersymmetric particles from a proton beam-dump
  experiment.
\newblock {\em Phys. Lett.}, 121B:429, 1983.

\bibitem{ball}
R.~C. Ball et~al.
\newblock {\em Phys. Rev. Lett.}, 53:1314, 1984.

\bibitem{bebc}
WA66 Collaboration.
\newblock Bounds on light gluinos from the bebc beam-dump experiment.
\newblock {\em Phys. Lett.}, 160B:212, 1985.

\bibitem{akesson}
T.~Akesson et~al.
\newblock {\em Z. Phys. C}, 52:219, 1991.

\bibitem{f:55}
G.~R. Farrar.
\newblock {\em Phys. Rev. Lett.}, 55:895, 1985.

\bibitem{f:24}
G.~R. Farrar and P.~Fayet.
\newblock {\em Phys. Lett.}, 79B:442--446, 1978.

\bibitem{ua1}
UA1 Collaboration.
\newblock {\em Phys. Lett.}, 198B:261, 1987.

\bibitem{franco}
E.~Franco.
\newblock {\em Phys. Lett.}, 124B:271, 1983.

\bibitem{hk:taugluino}
H.~Haber and G.~Kane.
\newblock {\em Nucl. Phys.}, 232B:333, 1984.

\bibitem{muller}
R.~Muller et~al.
\newblock {\em Science}, 196:521, 1977.

\bibitem{bei}
V.~S. Berezinsky, J.~Ellis, and B.~L. Ioffe.
\newblock {\em Phys. Lett.}, 172B:423, 1986.

\bibitem{bi}
V.~S. Berezinsky and B.~L. Ioffe.
\newblock Technical Report ITEP-127, ITEP preprints, 1985.

\bibitem{ov}
L.B. Okun and M.B. Voloshin.
\newblock {\em Sov. J. Nucl.Phys.}, 43:495, 1986.

\bibitem{bhk}
R.~Barnett et~al.
\newblock {\em Nucl. Phys.}, B267:625, 1986.

\bibitem{f:9}
G.~R. Farrar.
\newblock {\em Nucl. Phys.}, B77:429--442, 1974.

\bibitem{f:32}
G.~R. Farrar.
\newblock {\em Proceedings of the Cornell $Z^0$ Theory Workshop, Feb. 6-8,
  1981}, 1981.

\bibitem{jk}
M.~Jezabek and J.~Kuhn.
\newblock {\em Phys. Lett.}, B301:109, 1993.

\bibitem{f:82}
G.~R. Farrar.
\newblock {\em Phys. Lett.}, 265B:395, 1991.

\bibitem{m-ts}
R.~Munoz-Tapia and W.~Stirling.
\newblock {\em Phys. Rev.}, D49:3763, 1994.

\bibitem{opal4jet}
The~Opal Cllaboration.
\newblock Technical Report CERN-PPE/94-135, CERN, 1994.

\bibitem{enr}
J.~Ellis, D.~Nanopoulos, and D.~Ross.
\newblock {\em Phys. Lett.}, B305:375, 1993.

\bibitem{RtauRZ}
M.~Schmelling and R.~St.Denis.
\newblock {\em Phys. Lett.}, 329B:393, 1994.

\bibitem{bnp}
E.~Braaten, S.~Narison, and A.~Pich.
\newblock {\em Nucl. Phys.}, B373:581, 1992.

\bibitem{altarelli:hanoi}
G.~Altarelli.
\newblock Technical Report CERN-TH.7246/94, CERN, Proceedings of the 1993
  Rencontres du Vietnam, 1994.

\bibitem{clavelli}
L.~Clavelli.
\newblock {\em Phys. Rev.}, D45:3276, 1992.

\bibitem{clavetal}
L.~Clavelli et~al.
\newblock {\em Phys. Rev.}, D47:1973, 1993.

\bibitem{cs}
C.~E. Carlson and M.~Sher.
\newblock Photoproduction of very light gluinos.
\newblock {\em Phys. Rev. Lett.}, 72:2686, 1994.

\bibitem{bkn}
T.~Banks, D.~Kaplan, and A.~Nelson.
\newblock {\em Phys. Rev.}, D49:779, 1994.

\end{thebibliography}

\begin{figure}
\epsfxsize=\hsize
\epsffile{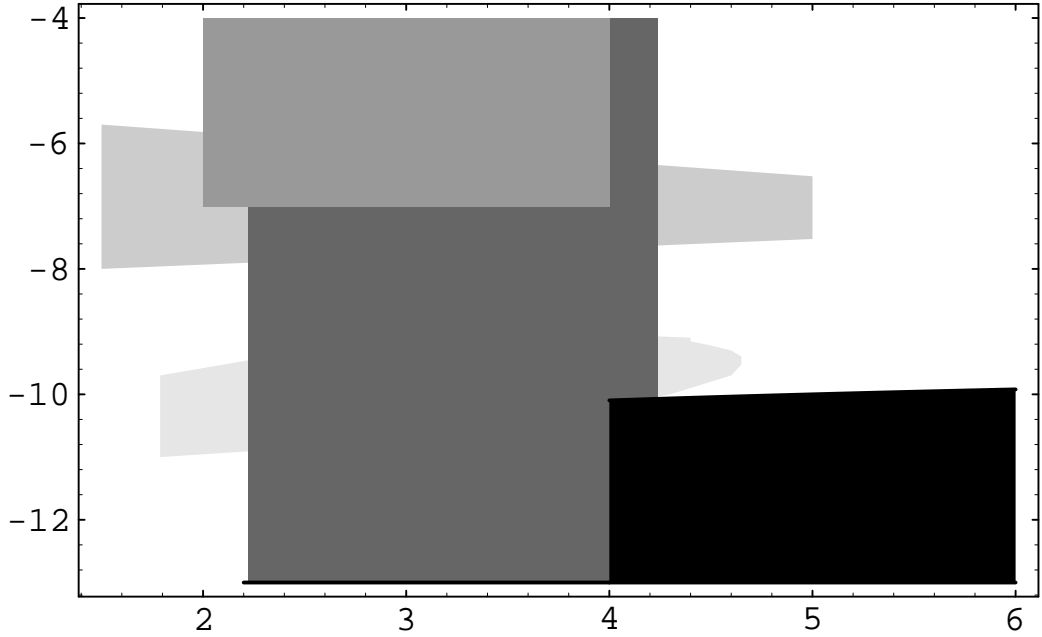}
\caption{Experimentally excluded regions of $m(R^0)$ and
$\tau_{\tilde{g}}$.  Horizontal axis is $m(R^0)$ in GeV beginning at 1.5
GeV;  vertical axis is $Log_{10}$ of the lifetime in sec.  A massless
gluino would lead to $m(R^0) \sim 1.4 \pm .4$ GeV.  ARGUS and
Bernstein et al give the lightest and next-to-lightest regions (lower
and upper elongated shapes), respectively.  CUSB gives the
next-to-darkest block; its excluded region extends over all lifetimes.
Gustafson et al gives the smaller (mid-darkness) block in the upper
portion of the figure; it extends to infinite lifetime.  UA1 gives the
darkest block in the lower right corner; it extends to higher masses
and shorter lifetimes not shown on the figure.}
\label{ltgl}
\end{figure}

\end{document}